# On Supporting IP Routing in the Next Generation of Mobile Systems

Hamed Hellaoui, Matti Laitila, Markus Isomäki, and Hua Chao

*Abstract*—The upcoming generation of mobile telecommunication systems is expected to support new use cases, where the mobile network serves one or more IP subnetworks located behind the User Equipment (UEs). This would create new challenges for the mobile system to efficiently serve such behind-UE subnetworks, as the latter are commonly not visible to the mobile system. In 3GPP, there have been works on Time-Sensitive Networking (TSN) and Deterministic Networking (DetNet), where the 5G System (5GS) is considered as a bridge or a DetNet node. In order to efficiently serve behind-UE IP subnetworks, we foresee the need for a further generalization where the mobile system (5GS and beyond) acts as a set of IP routers with more generic capabilities. In this article, we introduce the concept of Mobile System Router (MS-Router) which aims to provide a reference architectural design to enable the support of IP routing in the next generation of mobile telecommunication systems. The concept models a mobile system as an IP router per User Plane granularity. Each MS-Router implements an IP routing protocol, exchanges routing messages with the external routers and constructs a routing table, enabling the mobile system to dynamically learn the topology of the IP subnetwork behind the UEs and Data Network. The learned topology is translated into User Plane configuration to serve the IP subnetworks in an optimal way. The article also advocates different approaches where routing protocols can be implemented in the mobile system.

## I. INTRODUCTION

AS the 5th generation of telecommunication technology (5G) is reaching the final stages in terms of research and standardization, the focus has now turned to the next evolutionary step, also known as 5G-Advanced [1]. The latter will bring a new level of enhanced capabilities and enable a wider set of advanced use cases for verticals. 5G-Advanced will also pave the way for 6G as the upcoming generation of mobile telecommunication technology [2]. In this regard, one of the target use cases is the support of communication beyond the traditional User Equipment (UEs). Indeed, in different industrial applications, the mobile network is expected to serve one or more IP subnetworks located behind the UEs. Cooperating mobile robots and smart factory assembly are some examples, just to name a few [3], [4]. This would create new challenges for the mobile system to efficiently serve such behind-UE subnetworks, as the latter are commonly not visible to the mobile system. This would therefore require enhanced capability to be supported by the next generation of mobile systems.

H. Hellaoui, M. Laitila, and M. Isomäki are with Nokia, Espoo, Finland (e-mail: firstname.lastname@nokia.com).
H. Chao is with Nokia Shanghai Bell, Shanghai, China (e-mail: hua.chao@nokia-sbell.com).

Initiatives to support behind-UE subnetworks have already started in the current 5G specifications. 3rd Generation Partnership Project (3GPP) has introduced the support of framed routing and IPv6 delegation in 5G, so that the mobile system can know that a range of IPv4 addresses or IPv6 prefixes, assigned to hosts behind a UE, is reachable over a Protocol Data Unit (PDU) session [5]. This enables the network to route user traffic to these hosts. However, the mobile system still lacks the visibility of the underlying subnetwork topology, which limits routing user traffic in an optimal way. For example, in case that a host is reachable via two UEs, the mobile system cannot determine the UE over which the routing cost to this host would be minimal. On the other hand, 3GPP has also specified the support of Time-Sensitive Networking (TSN) and Deterministic Networking (DetNet), where the 5G System (5GS) is considered as a bridge or a DetNet node [6], [7]. These concepts also enable the support of behind-UE subnetworks, where a centralized controller has a complete view of the network and is able to configure the bridges/nodes. However, these two approaches are very specific and require dedicated environments. In particular, 5G-TSN necessitates TSN support in the system behind the UE, while 5G-DetNet requires DetNet support in this system. As many deployments rely on IP routing, these concepts have limited applicability.

In IP subnetworks, an IP router can run a routing protocol that determines the next node to forward user traffic to, so it can be routed from a source to a destination in an optimal way. As an example, the routing protocol "Open Shortest Path First" (OSPF) [8] is based on the Shortest Path First algorithm to determine the next node, where each router in the same area maintains an identical link-state database that describes the topology of the area. IP routing is widely employed to transport data over networks, and can be used by physical routers as well as by virtual routers (i.e., software routers deployed on the top of a generic hardware). With regard to a mobile system, IP routing is used both behind the UEs and the Data Network (DN).

In this article, we introduce the concept of Mobile System Router (MS-Router). This concept can be seen as a further generalization of the ongoing activity in 3GPP to support vertical requirements in terms of ensuring efficient connectivity to behind-UE subnetworks. The proposed MS-Router concept enables a mobile system to operate as a set of IP routers, bridging the gap between the IP subnetwork behind the UEs and the IP subnetworks behind the DN. In order to advance such a concept, this article addresses several challenging questions, which include the following:
- how to model an IP router in a mobile system and what is its granularity?

- at what level of the mobile system is the IP routing protocol implemented?
- how to convey routing messages from the external routers to the mobile system entity where the routing protocol is implemented?
- how to translate the learned topology into configuration to efficiently route user traffic?

The rest of this article is organized in the following fashion. Section II introduces some use case considerations for the next generation of mobile systems. These will thereafter be translated, in Section III, into an analysis on the need to support IP routing protocols in mobile systems. To this end, the proposed concept of Mobile System Router is introduced in Section IV. The future research directions are discussed in Section V. This article concludes in Section VI.

## II. USE CASE CONSIDERATION FOR THE NEXT GENERATION OF MOBILE SYSTEMS

One of the target use cases of the next generation of mobile systems is the support of communication beyond traditional UEs. For instance, in a smart factory scenario, the underlying tasks surpass the capabilities of a single robot and require the cooperation of multiple robots. These robots form a cluster and utilize local communication to establish a subnetwork. The latter is used for coordinating and controlling their actions, exchanging raw/processed data, exchanging knowledge about the environment, etc. This subnetwork's communication relies solely on IP routing, lacking support for TSN or DetNet technologies. As a result, 5G-TSN and 5G-DetNet functionalities are not applicable in this context. The robots are also connected to the mobile network to ensure connectivity to a local/central data center, or to enable remote communication and monitoring (e.g., software deployed in clusters located in a DN). In this regard, the mobile network is therefore serving an entire subnetwork rather than a simple UE [3], [9]. Fig. 1 provides an illustration of this use case.

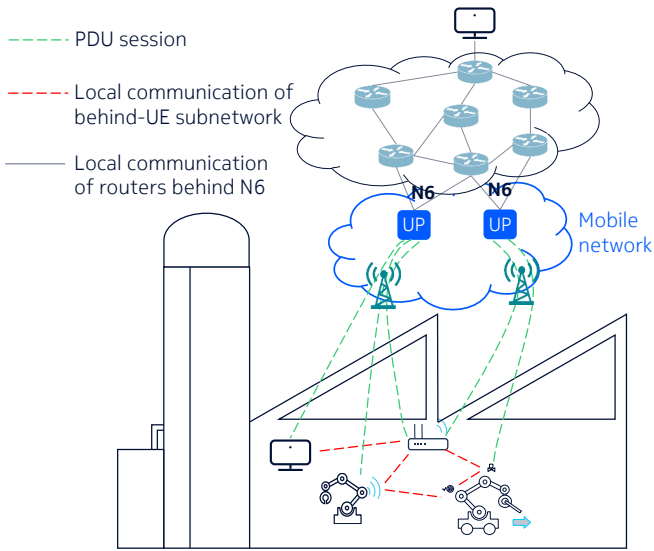

Fig. 1: A use case example.

As can be seen from Fig. 1, the communication between a host behind N6 interfaces and a device behind UEs can be performed via multiple routes. Consequently, communication through these diverse routes can exhibit variations and reachability issues due to factors such as link state changes, route congestion, or disconnections of battery-powered devices. This affects different segments, including the behind-UE subnetwork and the DN subnetwork. In order to ensure efficient communication in such scenarios, the following requirements need to be considered:

- On the downlink direction, it is not enough to only optimize the path between a host behind the N6 interface to the N6-routers (routers connected to the mobile system via the N6 interfaces). Indeed, this path should be selected so that the routing metrics between N6-routers, User Plane (UP) functions, UE-routers (routers connected to the mobile system via the PDU sessions), and also the behind-UE subnetworks are also considered.
- As for the uplink direction, selecting a routing path based only on the metrics available in the behind-UE subnetwork would not lead to optimal routing. Instead, the selected path should also consider the metrics between the UE-routers, UP functions, N6-routers, and the DN subnetwork.

It is to highlight that the above requirements cannot be ensured by the current specifications of 5G, which call for new approaches to support such scenarios.

## III. WHY SUPPORTING IP ROUTING IN MOBILE SYSTEMS?

As mentioned earlier, initiatives to support behind-UE subnetworks have already started in the current specifications of 5G. However, solutions such as TSN and DetNet (where 5GS is considered as a bridge or a DetNet node) do not enable the mobile system to have a visibility on the subnetwork topology behind the UEs. In this regard, we foresee a further generalization where the mobile system acts as a set of IP routers that can run IP routing protocols. Indeed, enabling such a concept would bring multiple benefits, including the following:

- By acting as an IP router, the mobile system can have complete visibility on the subnetwork topology behind the UE-routers as well as on the topology behind the N6-routers. This could include the link state between the routers as well as their IP addresses. The mobile system, as a router, would therefore be able to know if a device in a subnetwork behind the UE is active or down, in addition to the routing cost via this device. Moreover, by running a dynamic IP routing protocol, the mobile network can continuously have a visibility about these subnetworks. Such a dynamic visibility becomes very important as the subnetworks (behind UE-routers or N6-routers) could be subject to different changes (e.g., link state changes, congestion of routes, disconnection of battery-powered devices, etc.).
- Based on the learned topology, the mobile network can perform optimal decisions when routing user traffic to a behind-UE device, or to a host behind N6-routers. For instance, when such a device is reachable via two UE-routers, the mobile network can determine the UE-router over which the routing cost would be minimal. Such an optimal decision would not be possible without the knowledge about the underlying subnetwork topology. Indeed, by running a routing protocol such as OSPF, the mobile network can determine the shortest path to a given device/host residing in the behind-UE subnetwork or after the N6-routers. Such an optimal routing decision cannot be performed by the current 5G specifications.

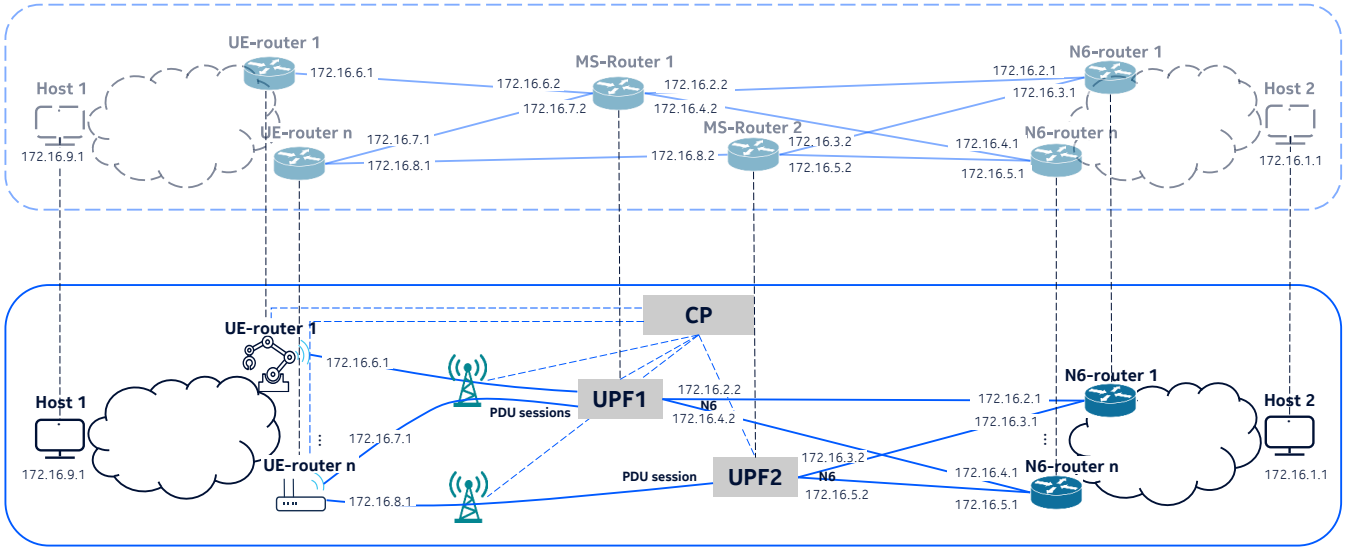

Fig. 2: Principle of a Mobile System Router.

- Furthermore, the external routers (in the behind-UE subnetwork as well as those behind N6-routers) would also benefit from the support of IP routing by the mobile system. Indeed, routers in a behind-UE subnetwork can learn the optimal path to forward uplink traffic via the mobile system, which is perceived as a router. Similarly, routers in the subnetwork behind the N6 interface can also learn the optimal route to forward downlink traffic to a behind-UE device via the mobile system.

The next section describes the proposed concept that enables the support of IP routing in 5G and beyond mobile systems.

## IV. TOWARDS SUPPORTING IP ROUTING IN 5G AND BEYOND MOBILE SYSTEMS

This section introduces the proposed concept to enable the support of IP routing protocols in 5G and beyond mobile systems. This is achieved by advancing the concept of MS-Router, which can be seen as a further generalization of some already standardized principles, such as TSN and DetNet. To this end, we first introduce the principle and the definition of an MS-Router. Thereafter, we further detail the possible implementation options. Furthermore, we also provide a discussion on the capabilities and advantages offered by the proposed MS-Router concept.

### A. Principle of a Mobile System Router

Fig. 2 illustrates the principle of MS-Router, where the mobile system can be seen as a set of IP routers that implement IP routing protocols. Similar to TSN and DetNet, the granularity of an MS-Router is per UP Function (UPF), which is translated into one MS-Router per UPF. Indeed, user traffic is processed/routed at the level of UPF, and scoping the granularity of an MS-Router to a UPF-level would reflect the appropriate routing performed by a mobile system. An MS-Router also leverages the Control Plane (CP) functions and procedures of the mobile system in order to operate appropriately and ensure maximum backward compatibility. This is further detailed in the next subsection. As illustrated in Fig. 2, the mobile system at the bottom of the figure has two UPFs (UPF1 and UPF2) and can be seen as two routers (resp. MS-Router 1 and MS-Router 2) as highlighted in the top of the figure. Each MS-Router can communicate with the routers reachable via the N6 interfaces, as well as with UE-routers reachable via the PDU sessions. The interfaces of each MS-Router become *i)* the N6 interfaces and *ii)* the PDU sessions.

Operating a routing protocol requires that the latter is aware of the interfaces, their identifiers, and their IP addresses. In the case of MS-Router, the identifiers of N6 interfaces and PDU sessions are used to identify the router interfaces. These identifiers are mapped to standard interface names, so they can be understood by the routing protocol. On the other hand, while N6 interfaces are already associated with IP addresses, PDU sessions are not. The MS-Router needs therefore to reserve an IP address, for each PDU session interface, in the same subnet as the IP address allocated to the UE that is using this PDU session. As illustrated in Fig. 2, the two MS-Routers highlighted in the top of the figure use the same IP addresses as their corresponding N6 interfaces, while they use new IP addresses for their PDU session interfaces in the same subnets as the corresponding ones considered for the UEs. Note that the same principle is applicable for a UE having several PDU sessions anchored to the same UPF (i.e., each of these PDU sessions will be considered as an interface for this MS-Router, which will reserve an IP address for each interface in the same subnet as the corresponding one considered for the UE). Note also that Fig. 2 depicts the MS-Router concept using private IP addresses class B for illustrative purposes. While this example uses private addresses, other scenarios are possible, such as the host behind the N6 interfaces utilizing public IP addresses. It's important to highlight that using private IP addresses for router interfaces is a common practice. A summary of MS-Router features is provided in Table I.

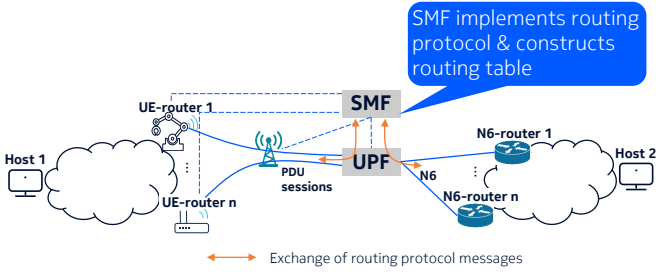
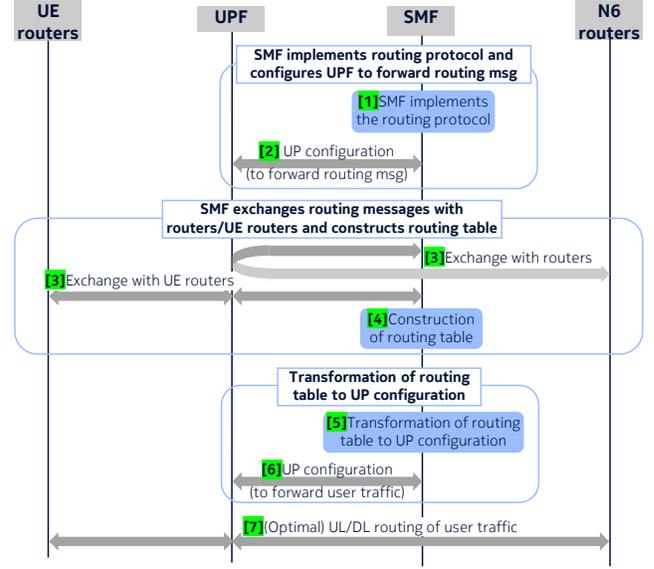

(a) Architectural view

(b) Sequence flow

Fig. 3: Approach 1 (routing protocol implemented by SMF).

TABLE I: Summary of MS-Router features.

| Feature | Description |
|---|---|
| Granularity | Per User Plane Function |
| Interfaces | N6 interfaces & PDU sessions |
| IDs of interfaces | - IDs of N6 interfaces<br>- IDs of PDU sessions<br>(all mapped to standard interface names) |
| IP addresses of interfaces | - IP addresses of N6 interfaces<br>- MS-Router reserves IP address for each PDU session interface in the same subnet as the corresponding one considered for the UE |

### B. Implementation options

This subsection tackles the implementation options of a mobile system router. As mentioned earlier, we propose that a mobile system can become an IP router per a UPF granularity (where the router interfaces become the N6 interfaces and the PDU sessions). We advocate two implementation options, which are distinguished according to the level in which the IP routing protocol is implemented. These two options are CP-based implementation and UP-based implementation.

*1) Approach 1 - Implementation of the IP routing protocol in the CP:* In this approach, the routing protocol is implemented in the CP of the mobile system. Fig. 3 (a) shows an architectural view of this approach, where the routing protocol is implemented in the Session Management Function (SMF). The latter will therefore exchange routing information with the external routers (N6-routers and UE-routers) and construct the routing table. The choice for the SMF to implement the routing protocol is supported by the fact that the SMF is the only CP function in 5GS that supports the exchange of UP traffic with User Plane functions (via a General packet radio service Tunnelling Protocol -GTP- tunnel over N4) [10]. On the other hand, the UPF needs to be configured to forward routing information between the SMF and the N6-routers, and also between the SMF and UE-routers. This would ease backward compatibility with the existing specifications.

A generic sequence flow for approach 1 is illustrated in Fig. 3 (b). As a pre-requirement, the SMF implements the routing protocol (**step 1** of Fig. 3 (b)). This means that the SMF knows the ID and the IP address of each MS-Router interface (i.e., N6 interfaces and PDU session interfaces). Considering the N6 interfaces, the SMF can know the ID and IP address of each interface from the UPF. As for PDU session interfaces, an IP address will be reserved for each PDU session as mentioned earlier.

As the routing protocol is implemented in the SMF, the latter needs to configure the UPF to forward IP routing messages between the SMF and the external routers (**step 2** of Fig. 3 (b)). This concerns the IP routers reachable via the N6 interfaces and also the UE-routers reachable via the PDU session interfaces, as illustrated in Fig. 3 (a). To this end, GTP tunnels (with TEIDs -Tunnel Endpoint Identifiers- corresponding to the interfaces) between the SMF and the UPF can be used to enable the SMF to exchange routing information with the external routers via the UPF. Therefore, at this step the SMF configures the UPF to transparently forward IP routing messages between the interfaces and the GTP tunnels with the corresponding TEIDs.

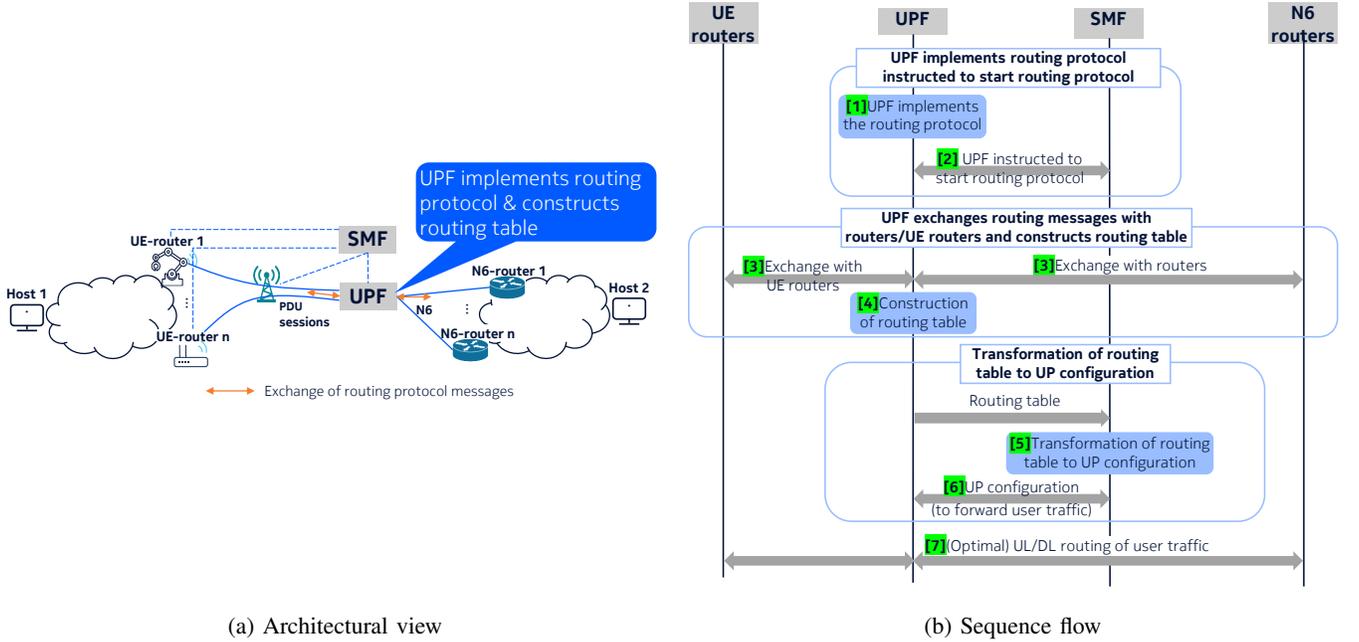

Fig. 4: Approach 2 (routing protocol implemented by UPF).

In **step 3** of Fig. 3 (b), the SMF exchanges IP routing messages with the external routers over the configured UPF. In this regard,

- when SMF aims to send an IP routing message via an N6 interface (resp. via a PDU session interface), it uses the IP address of the corresponding N6 interface (resp. the reserved IP address of the corresponding PDU session) as a source IP address in the packet header;
- when SMF aims to send an IP routing message via an interface, it uses the GTP tunnel with the corresponding TEID;
- the UPF is therefore forwarding IP routing messages between an interface and the GTP tunnel with the corresponding TEID;
- when SMF receives an IP routing message, it can know the corresponding interface based on the used GTP tunnel and the associated TEID.

Based on the IP routing messages exchanged between the SMF and the external routers, the SMF constructs a routing table (**step 4** of Fig. 3 (b)). How the SMF constructs the routing table may depend on the IP routing protocol supported by the SMF. Details are out of the scope of this article. Without losing generalization, Table II shows an illustration of a routing table constructed by the SMF. This illustrative table is based on Fig. 2 and reflects an example routing table constructed by MS-Router 1. As we can see, the mobile system (by its MS-Router 1) can become aware that both Host 1 and Host 2 are reachable via several paths with different metrics. This would allow taking effective decisions in terms of routing. Note that the routing protocol run by the SMF could be dynamic and the SMF would dynamically get the updates about the paths.

TABLE II: An example of a routing table constructed by the mobile system (MS-Router 1).

| Destination | Next Hop | Destination interface | Metric |
|---|---|---|---|
| 172.16.1.1 (Host 2) | 172.16.2.1 (N6-router 1) | ID of N6 interface 1 | 338 |
| 172.16.1.1 (Host 2) | 172.16.4.1 (N6-router n) | ID of N6 interface 2 | 292 |
| 172.16.9.1 (Host 1) | 172.16.6.1 (UE-router 1) | ID of PDU session 1 | 68 |
| 172.16.9.1 (Host 1) | 172.16.7.1 (UE-router n) | ID of PDU session n | 258 |
| ... | ... | ... | ... |

Once a routing table is learned, the SMF transforms it into configuration to route user data (**step 5** of Fig. 3 (b)). Considering the previous example, the SMF can know that the shortest path to Host 1 would be via UE-router 1 and the shortest path to Host 2 would be via N6-router n, and would consequently generate the associated UP configuration. Note that these configurations in 5G are based on Packet Forwarding Control Protocol (PFCP). These configurations will thereafter be installed to the UPF as reflected by **step 6** of Fig. 3 (b). At the end of this process, optimal routing can be performed by the mobile system in both uplink and downlink direction (**step 7** of Fig. 3 (b)). Note that as an MS-Router runs a dynamic routing protocol, steps 3, 4, 5, and 6 of Fig. 3 (b) are continuously executed, enabling to dynamically learn the topology and to transform it into optimal routing configuration, even in the face of changing network conditions.

*2) Approach 2 - Implementation of the IP routing protocol in the UP:* Unlink the previous approach, the routing protocol in approach 2 is implemented in the UP. An architectural view of this approach is illustrated in Fig. 4 (a), where the UPF is the entity implementing the routing protocol. This implies that the UPF will exchange routing information with the external routers (N6-routers and UE-routers) and construct the routing table.

Fig. 4 (b) depicts a generic sequence flow for approach 2. As reflected by **step 1** of Fig. 4 (b), the UPF implements the routing protocol as a pre-requirement. Consequently, the UPF knows the ID and the IP address of each MS-Router interface (i.e., the IP address of each N6 interface and the reserved IP address for each PDU session interface as discussed earlier). This also means that the UPF will terminate IP routing messages received from the external routers (UPF will not forward those messages). Unlike the previous approach, there is no need to rely on GTP tunnels between the SMF and UPF, as IP routing messages will be processed locally by the UPF.

Operating the routing protocol at the UPF could be triggered by an indication from the CP, as reflected by **step 2** of Fig. 4 (b). Thereafter, the UPF can exchange IP routing messages with the external routers (**step 3** of Fig. 4 (b)). Similar to the approach 1, when UPF aims to send an IP routing message via a PDU session interface, it uses the reserved IP address for this interface as a source IP address in the packet header. The UPF will therefore be able to construct a routing table (**step 4** of Fig. 4 (b)) based on the IP routing messages exchanged with external routers. The constructed routing table could also be illustrated by Table II.

Similar to the previous approach, the constructed routing table is converted into configurations to route user traffic. To this end, the UPF sends the routing table to the CP, which then translates it into UP configurations that will be installed at the UPF (**step 5** and **step 6** of Fig. 4 (b)). This would lead to optimal routing of user traffic by the mobile system in both uplink and downlink directions (**step 7** of Fig. 4 (b)). Note also that steps 3, 4, 5, and 6 of Fig. 4 (b) are continuously executed, enabling the dynamic topology learning and the generation of optimal routing configurations, even in the presence of evolving network conditions.

*C. Discussion*

The proposed MS-Router concept empowers the mobile system to function as a group of IP routers. These routers can execute IP routing protocols, exchange routing information with external routers, and build routing tables. Consequently, an MS-Router can dynamically learn the IP subnetwork topology behind UEs and N6 interfaces. This topology information is then transformed into UP configurations, enabling optimal traffic routing on both uplink and downlink. This goes beyond the simple use of framed routing and IPv6 delegation, where the mobile system can only know that a range of IPv4 addresses or IPv6 prefixes is reachable over a PDU session. On the other hand, the adoption of TSN and DetNet enables the support of behind-UE subnetworks. However, these two concepts are very specific and require a dedicated environment. While 5G-TSN operates within a TSN environment and 5G-DetNet functions within a DetNet environment, the proposed MS-Router concept focuses on IP subnetworks that utilize IP routing protocols.

As introduced in the previous subsection, two implementation options of an MS-Router can be considered. The key distinction between the two approaches lies in the implementation level of the IP routing protocol. In approach 1, the routing table is built within the CP, while the UP handles forwarding routing protocol messages between the CP and UE-routers/N6-routers. In contrast, approach 2 implements the IP routing protocol directly in the UP. This difference implies that approach 2 benefits from faster routing table construction due to its implementation within the UP. This advantage stems from the fact that approach 1 involves forwarding routing protocol messages through the UP, adding an extra hop compared to approach 2. On the other hand, approach 1 offers faster transformation of the routing table into UP configuration, as the table is already established in the CP.

The proposed solution exhibits a key strength in its protocol-agnostic nature. It can seamlessly accommodate various routing protocols, including OSPF, Routing Information Protocol (RIP), Intermediate System to Intermediate System (IS-IS), and Border Gateway Protocol (BGP). This flexibility is highly advantageous from a standardization standpoint, enabling operators to choose the most suitable routing protocol for their specific needs. The selection of a routing protocol for the MS-Router is primarily influenced by the protocols employed by the N6-routers and UE-routers, as well as the desired network objectives. For example, if both N6-routers and UE-routers utilize OSPF and the goal is to ensure consistent topology awareness across all routers, the MS-Router could also adopt OSPF within the same area.

As an MS-Router runs a routing protocol, it adheres to the protocol's defined processes. These processes encompass various operations such as neighbor discovery, neighbor maintenance, and cost calculation. For example, OSPF utilizes five packet types, with type 1 being the "Hello message" used for neighbor discovery and maintenance. Consequently, an MS-Router running OSPF will discover its neighbors by sending this message. Similarly, N6-routers and UE-routers will also discover MS-routers as part of the routing protocol processes. OSPF defines other packet types related to link state information, enabling the sharing of topological views and the handling of failures. Since OSPF relies on the "Shortest Path First" algorithm for cost calculation, an MS-Router running this protocol will be able to determine the shortest path to target nodes located behind UEs or N6 interfaces.

In practice, multiple scenarios exist regarding IP address allocation. While UEs are typically assigned private IP addresses, it is technically feasible to allocate public IP addresses to them (indeed, many mobile operators offer this option for an additional fee). Similarly, devices behind UEs are usually assigned private IP addresses, and various techniques like address translation by UEs, framed routing, or IPv6 prefix delegation can be employed. The MS-Router concept is designed to function within these scenarios. It operates by running a routing protocol and establishing connections with UE-routers. For each PDU session interface, the MS-Router utilizes an IP address within the same subnet as the corresponding one considered for the UE, enabling a proper communication with the UE-routers.

## V. Future Research Directions

The consideration of the proposed MS-Router concept would enable a mobile system to act as IP routers, learn the network topology behind the UE-routers and the N6-routers, and route user traffic over the optimal routes. However, some research challenges, with standard implications, need to be addressed.

Firstly, the behavior of an MS-Router and the way the implemented routing protocol is configured should follow the policy of the mobile system. In the current specifications, policy and charging control framework is governed by a dedicated entity [11]. The integration of the MS-Router concept with the mobile system would require standardizing the interaction with policy and control function to enable the appropriate configuration of the underlying routing protocol. Furthermore, an MS-Router will exchange routing information with UE-routers and/or N6-routers. This could include some internal information, such as link state of the PDU session interfaces, neighbor list, etc. The exchange of such information should also follow the policy of the mobile system.

Considering the proposed MS-Router concept, the N6-routers can learn the topology of the network behind the UE-routers as part of the learning process of the routing protocol. This mainly depends on the used routing protocol and the way it is configured. For instance, by considering the routing protocol OSPF with one area for the N6-routers, the MS-Routers, and the UE-routers, each of these routers would have the same knowledge of the topology as part of the learning process of OSPF. However, 5GS already considers advertising framed routing to the N6-routers, so that the latter can know routes/prefixes reachable via a UPF [12]. Such advertisement would therefore become unnecessary (duplication) in such a situation, which would require introducing new mechanisms to control the advertisement of framed routing when the N6-routers can discover the network topology behind the UE-routers as part of the MS-Router concept.

When the routing protocol is implemented in the control plane (approach 1), the exchanges of IP routing messages between the CP and the UE-routers can either be done via the UP (in this case, the UPF needs to be configured to forward the IP routing messages between the CP and the UE-routers) or directly via Non-Access Stratum (NAS) signalling. These represent two different options that their advantages and drawbacks should be investigated.

Enabling the communication of an MS-Router with a UE-router would require reserving an IP address for each PDU session interface in the same subnet as the corresponding one considered for this UE (so that the routing protocol implemented by the MS-Router can use it as a source IP address in the packet header when communicating with UE via this interface). In the current specification, a UE can get its IP address by different means, including through the use of the Dynamic Host Configuration Protocol (DHCP) or NAS signalling. In the two options, standard specifications are required to enable the MS-Router to know and reserve such IP addresses.

On the other hand, as a PDU session becomes an interface for the MS-Router, the latter needs to consider the associated changes that a PDU session may undergo. Indeed, a PDU session can be associated with different management operations, including session establishment, session modification and session release. Furthermore, unlike traditional routers, UE-routers could be mobile, which may invoke some session management operations.

We have advanced in this article two approaches for implementing a mobile system router. While in the first approach the routing protocol is implemented in the control plane, the second approach is based on implementing the routing protocol in the user plane. Exploring the possibility of combining the two approaches is an open question.

## VI. Conclusion

The emerging applications in which the mobile system needs to efficiently serve one or more IP subnetworks behind the UEs create new challenges and call for new approaches. This article introduced the concept of MS-Router, that enables a mobile system to act as a set of IP routers. The concept allows to bridge the IP subnetworks behind the UEs with those behind the N6 interfaces. Each MS-Router is modeled per a User Plane granularity, and exchanges routing information with external routers (UE-routers and N6-routers). The mobile system would therefore be able to have a visibility on these subnetworks and to take optimal decisions for routing user traffic. Furthermore, routers in the behind-UE subnetwork and those behind N6 interface will also perceive the mobile system as (a set of) routers and learn optimal routes via the mobile system. Two implementation approaches have been discussed in the article (i.e., CP-based implementation and UP-based implementation). The proposed concept can be considered as a further generalization of the ongoing activities in 3GPP to support behind-UE subnetworks.